%% file: paper.tex
\documentclass[preprintnumbers,amsmath,amssymb]{revtex4}

\usepackage{graphicx}
\usepackage{dcolumn}
\usepackage{bm}

\begin{document}

\title{Baryon-Meson Loop Effects on the Spectrum of Non Strange Baryons}

\author{Danielle Morel}
 \email{dmorel@descartes.physics.fsu.edu}
\author{Simon Capstick}%
 \email{capstick@csit.fsu.edu}
\affiliation{%
Department of Physics, Florida State University, 
Tallahassee, FL 32306-4350, USA}%

\date{\today}

\begin{abstract}
Corrections to the masses of baryons from baryon-meson loops can
induce splittings between baryons which are comparable to those
arising from the residual interactions between the quarks. These
corrections are calculated using a pair-creation model to give the
momentum-dependent vertices, and a model which includes configuration
mixing to describe the wave functions of the baryons. A large set of
baryon-meson intermediate states are employed, with all allowed
SU(3)$_{\rm f}$ combinations, and excitations of the intermediate
baryon states up to and including the second band of negative-parity
excited states. Roughly half of the splitting between the nucleon and
Delta ground states arises from loop effects. The effects of such
loops on the spectrum of negative-parity excited states are examined,
and the resulting splittings are sensitive to configuration mixing
caused by the residual interactions between the quarks. With
reduced-strength one-gluon-exchange interactions between the quarks
fit to the Delta-nucleon splitting, a comparison is made between model
masses and the bare masses required to fit the masses of the states
extracted from data analyses. This shows that it is necessary to also
adjust the string tension or the quark mass to fit the splitting
between the average bare masses of the ground states and
negative-parity excited states, and that spin-orbit effects are likely
to be important.
\end{abstract}

\pacs{12.40.Yx, 12.39.-x, 12.39.Pn, 12.39.Jh}
\maketitle
%
%
\def\slash#1{#1 \hskip -0.5em / }  
\def\rmb#1{{\bf #1}}
\def\lpmb#1{\mbox{\boldmath $#1$}}
\def\nn{\nonumber}
\def\>{\rangle}
\def\<{\langle}
\newcommand{\Eqs}[1]{Eqs.~(\protect\ref{bib#1})}
\newcommand{\Eq}[1]{Eq.~(\protect\ref{#1})}
\newcommand{\Fig}[1]{Fig.~\protect\ref{#1}}
\newcommand{\Figs}[1]{Figs.~\protect\ref{#1}}
\newcommand{\Sec}[1]{Sec.~\protect\ref{#1}}
\newcommand{\Secs}[1]{Secs.~\protect\ref{#1}}
\newcommand{\Ref}[1]{Ref.~\protect\cite{#1}}
\newcommand{\Refs}[1]{Refs.~\protect\cite{#1}}
\newcommand{\Tab}[1]{Table~\protect\ref{#1}}
\newcommand{\sfrac}[2]{\mbox{$\textstyle \frac{#1}{#2}$}}
%
%
%
\section{\label{sec:Intro}Introduction}
%
In QCD there are $qqq(q\bar{q})$ configurations possible in baryons,
and these must have an effect on the constituent quark model, similar
to the effect of unquenching lattice QCD calculations. These effects
can be modeled by allowing baryons to include baryon-meson (B$^\prime$M)
intermediate states, which lead to baryon self energies and mixings of
baryons of the same quantum numbers. A calculation of these effects
requires a model of baryon-baryon-meson (BB$^\prime$M) vertices and
their momentum dependence. It is also necessary to have a model of the
spectrum and structure of baryon states, including states not seen in
analyses of experimental data, in order to provide wave functions for
calculating the vertices, and to know the thresholds associated with
intermediate states containing missing baryons.

Baryon self energies due to B$^\prime$M intermediate states and
B$^\prime$M decay widths can be found from the real and imaginary
parts of loop diagrams. The size of such self energies can be expected
to be comparable to baryon widths. For this reason, they cannot be
ignored when comparing the predictions of any quark model with the
results of analyses of experiments. Since the mass splittings between
states which result from differences in self energies are likely
similar to those that arise from the residual interactions between the
quarks (defined to be interactions which are present after taking into
account confinement), a self-consistent calculation of the spectrum
needs to adjust the residual interactions, and so the wave functions
of the states used to calculate the BB$^\prime$M vertices, to account
for these additional splittings.

\begin{figure}
\hspace{-0.3in}
\includegraphics[width=6.9cm,angle=0]{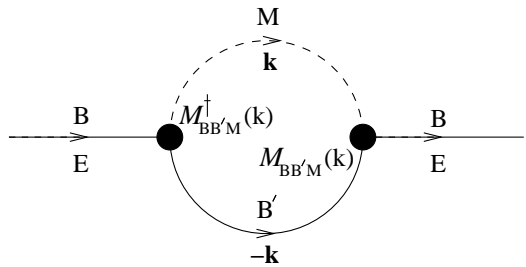}
\caption{\label{loop} Baryon-meson $B^\prime M$ loop contributing to
the energy-dependent self-energy $\Sigma_B(E)$ of a baryon $B$ with
bare mass $E$, evaluated in the rest frame of the initial baryon. The
intermediate baryon and meson are assigned physical masses
$m_B^\prime$ and $m_M$.}
\end{figure}

In time-ordered perturbation theory (TOPT), the contribution to the
self-energy of a baryon $B$ with bare energy $E$ from the baryon-meson
loop illustrated in Figure~\ref{loop} is
\begin{equation}
\label{SigmaBB'M(E)}
\Sigma^{\rm B}_{B^\prime M}(E)=\int d\lpmb{k} 
{{\cal M}^\dag_{\rm BB^\prime M}(k) {\cal M}_{\rm BB^\prime M}(k)
\over E-\sqrt{m_{\rm B'}^2+k^2}-\sqrt{m_{\rm M}^2+k^2}},
\end{equation}
where the calculation is carried out in the center-of-momentum frame
of the initial baryon. Note that the intermediate baryon and meson are
assigned their physical masses $m_{B^\prime}$ and $m_M$.  This
ensures~\cite{Silvestre-Brac:pw} that the poles due to decay
thresholds are in their correct positions~\cite{dressed-bare}.
%
%

The strong decay matrix element ${\cal M}_{BB^\prime M}(k)$ depends on
the loop momentum and the spin, flavor and spatial structure of the
hadrons involved, and plays the role of a form factor. When the
effects of confinement are included in the spatial structure of the
hadrons, the factor $|{\cal M}_{BB^\prime M}(k)|^2$ has the effect of
suppressing high-momentum contributions to the loop and rendering it
finite. Care has to be taken to evaluate the principal part of the
loop integral when it crosses a pole, present if the bare energy $E$
is sufficiently large to allow the decay $B\to B^\prime M$ to
proceed. The imaginary part of the loop integral is given by the
residue of this pole, and is related to the partial width for this
decay. A model of these strong-decay vertices can, therefore, be fit
to the physical decay rates, and can then be used to predict the
variation of the strong decay matrix element ${\cal M}_{BB^\prime
M}(k)$ away from the physical decay point.

The self energy of a baryon $B$ is then evaluated by adding the
contributions from all possible intermediate loops 
\begin{equation}
\label{SigmaB(E)}
\Sigma_{\rm B}(E)=\sum_{\rm B'M}\int d\lpmb{k} 
{{\cal M}^\dag_{\rm BB^\prime M}(k) {\cal M}_{\rm BB^\prime M}(k)
\over E-\sqrt{m_{\rm B'}^2+k^2}-\sqrt{m_{\rm M}^2+k^2}}
\end{equation}
It is crucial that this sum is over a set of intermediate states which
is large enough that the differences in the self energies (the self
energies themselves are not observables) do not change appreciably
with the inclusion of additional states. This is a non-trivial
requirement, especially when the intermediate state includes excited
baryons, since the baryon spectrum includes a large number of excited
states close in energy to the ground states. Not only does the sum
have to include (ground state) baryons and mesons of different flavors
and total quark spins, it also in principal should include spatial
excitations of both the baryons and mesons. For larger hadron $J$
values, more than one relative angular momentum of the intermediate
hadrons is possible. The resulting complexity has often led in
previous calculations to premature truncation of the sum in
Eq.~(\ref{SigmaB(E)}).

Ignoring, for the moment, spatial excitations of the intermediate
hadrons, it is possible to define a complete set of spin-flavor
symmetry-related intermediate states. Consider the effects of
baryon-meson intermediate states on the $\Delta$-$N$ mass splitting,
traditionally used to set the strength of the spin-dependent contact
interactions between the quarks. If these spin-dependent interactions
are turned off, it is still possible that these states have self
energies from B$^\prime$M loops which are different, and so cause a
splitting between the states. Assume for now that there are only
ground-state baryons and mesons, made up of $u$, $d$, and $s$ quarks
with the same mass, and that there are no interactions between the
quarks other than confinement. In this SU(6)-symmetry limit all
(ground state) baryons will have the same masses and wave functions,
and the same is true of mesons. All of the intermediate states
B$^\prime$M used to calculate baryon self energies will have the same
mass $m_{\rm B^\prime}+m_{\rm M}$, and the energy-dependent self
energies in Eq.~(\ref{SigmaB(E)}) will differ only because of the
flavor and spin structure of the strong decay matrix elements ${\cal
M}_{BB^\prime M}(k)$.

It should be true that in this limit all ground state baryon self
energies are the same, and this has been demonstrated to be true in
Ref.~\cite{PZ}, {\it if} the set of intermediate states includes a
complete set of spin-flavor [SU(6)] symmetry related allowed
combinations of the ground state octet and decuplet baryons, and
ground state pseudoscalar and vector mesons. In particular, the
ground-state $\Delta$ and nucleon states will be degenerate in this
limit only if non strange and strange pseudoscalar and vector mesons
are included in the intermediate states. This implies that
calculations of self energies which do not include vector mesons, for
example, do not start at this limit and so cannot be expected to
produce physically meaningful results away from it.

With the exception of the work of Ref.~\cite{BB}, previous
calculations of the self energies of ground state and negative-parity
excited baryons use baryon-meson intermediate states including only
baryon (spatial) ground
states~\cite{PZ,BHM,Silvestre-Brac:pw,Fujiwara}. However, the
importance of including spatially-excited baryon states has been
established in a calculation~\cite{BB} of the $\Delta$-$N$
splitting. Here the intermediate states are restricted to the set of
states ${\rm B}^\prime \pi$, with baryon states B$^\prime$ chosen from
a set of ground and excited $N$ and $\Delta$ states. The sum over
intermediate states is shown to converge to a stable result for the
$\Delta-N$ splitting only when excited states from the $N=0$ (ground),
$N=1$ (lowest-lying negative-parity excited), $N=2$ (positive-parity
excited), {\it and} $N=3$ (highly-excited negative-parity) bands of
states are included. Similar results for the convergence properties of
the $\rho$ and $\omega$ meson self energies are found in a calculation
of the effects of meson-meson intermediate states on meson
masses~\cite{GI}. This suggests that calculations of baryon self
energies which restrict the intermediate baryons to (spatial) ground
states cannot be expected to have converged. 

In principle it may also be necessary to check for convergence of a
sum over spatially-excited intermediate meson states. However, since
(i) orbital and radial excitations of mesons tend to be significantly
more massive than their corresponding ground states, and (ii)
introducing spatial excitation into the meson wave function generally
reduces the size of the BB$^\prime$M vertices, and (iii) the
multiplicity of spatially-excited meson states is much lower than that
of spatially-excited baryon states, it may be possible to ignore
spatially-excited mesons in the set of intermediate states. This is
the approach adopted here.

Interestingly, in the model of Ref.~\cite{BB}, the difference in the
pionic self energies of the odd-parity excited states and the ground
state converges too slowly to make definite conclusions. This may be
due to the choice of $BB^{\prime}M$ amplitudes, where
pions are emitted directly from the quarks with a (non relativistic)
pseudoscalar coupling, and an additional (somewhat hard) axial form
factor
\begin{equation}
F_\pi({\bf k}^2)=1/(1+{\bf k}^2/\Lambda_{\pi}^2),
\end{equation}
with $\Lambda_{\pi}=1275$ MeV corresponding to the mass of the $a_1$
meson. Since the loop integrals involve elementary intermediate pions,
a factor of $1/\omega_k$ is included, where $\omega_k=\sqrt{{\bf
k}^2+m_\pi^2}$ is the pion energy, from the normalization of the wave
function of the intermediate pion. Note that this factor is not
present in the pion center of mass wave function in non relativistic
models which treat it as a composite particle. Although the presence
of this factor has the effect of further suppression of high-momentum
contributions to the integral over the loop
momentum~\cite{Silvestre-Brac:pw}, the net result is an effective
pion-nucleon vertex which is probably too hard. The same is likely to
be true of the work of Ref.~\cite{BHM}, which also uses an
elementary-meson emission strong-decay model for the $BB^\prime M$
vertices. In subsequent models and the present work a more rapid
decrease of the vertex amplitudes with $k^2$ is shown to produce
better results for the mass shifts, and this can be attributed to an
effective size for the operator which creates a constituent
quark-antiquark pair~\cite{Silvestre-Brac:pw,GI}. The issue of the
poor convergence of the sum over intermediate excited baryons found in
Ref.~\cite{BB} of the self energies of the negative-parity excited
states will be resolved in the present work.

This illustrates the importance of the choice of model to describe the
strong vertex amplitudes ${\cal M}_{BB^\prime M}(k)$. In particular,
it is necessary to take into account the spatial structure of the
emitted mesons to avoid vertex amplitudes which do not sufficiently
suppress the contributions of the $B^\prime M$ loops at large relative
momentum between the two hadrons. At the same time, the vertex
amplitudes should contain information about the spatial structure of
the initial and intermediate hadron states. A popular phenomenological
strong-decay model based on the creation of a $q\bar{q}$ pair with
vacuum ($^3P_0$) quantum numbers and applied to a baryon and meson
strong decays~\cite{Le-Yetal} has been adopted in previous
calculations of loop effects~\cite{PZ,Silvestre-Brac:pw}. In the
calculations of Refs.~\cite{PZ,Silvestre-Brac:pw}, a single baryon
radius and meson radius and unmixed harmonic-oscillator wave functions
were used to describe the baryon states, which is equivalent to
assuming SU(6) symmetry in the wave functions. These approximations
are not made in the present work. A more sophisticated decay model,
similar to that developed for mesons in Ref.~\cite{ABS}, was adopted
in the work of Ref.~\cite{Fujiwara}. This work also uses
antisymmetrized $(3q)(q\bar{q})$ cluster-model wave functions composed
of simple harmonic oscillator wave functions and plane-wave relative
motion to describe the baryon-meson intermediate states.

An important goal of prior calculations of the mass splittings of
negative-parity excited baryon arising from self energies is a
possible resolution of the spin-orbit problem in baryons. In general,
spin-orbit effects are too large when calculated with one-gluon
exchange residual interactions with the strength required to cause all
of the contact splittings between ground states and negative-parity
excited states evident in the spectrum. The situation is different
when the residual interactions between the quarks are assumed to arise
from one-boson exchange, as there are no corresponding spin-orbit
interactions, but those arising from Thomas precession in the
confining potential should still be present and will not be negligibly
small~\cite{IsgurSO,GlozmanSO}. With the introduction of mass
splittings due to loop effects it is possible~\cite{PZ,BB} to use a
reduced-strength residual interaction (in one-gluon exchange models
this means a smaller value of $\alpha_s$ in the limit of low $Q^2$),
which will naturally reduce the size of the resulting spin-orbit
effects~\cite{PZ}.

A second possibility is that, in addition to reducing the strength of
the residual interactions required to fit the observed splittings, it
may happen that the self energies induce splittings between the bare
masses which resemble spin-orbit effects.
Ref.~\cite{Silvestre-Brac:pw} explores the possibility that any
spin-orbit splittings in the spectrum of low-lying negative-parity
excited $N$, $\Delta$, $\Lambda$ and $\Sigma$ states arises from the
effects of differences in the self energies. It may also be possible
that the spin-orbit splittings due to differences in the self energies
are opposite in sign to those expected from
one-gluon-exchange~\cite{BHM}. This intriguing possibility has been
explored in Ref.~\cite{Fujiwara}. The results show that it may be
possible to arrange a cancellation between spin-orbit splittings
arising from the interactions between the quarks and from loop
effects, and to describe the mixings and decay widths of these states
in the same model. Notable exceptions are the flavor singlet (lowest
lying) negative-parity $\Lambda$ states $\Lambda(1405)$ and
$\Lambda(1520)$ which are about 100 MeV too heavy, as in simple
three-quark models.

Conclusions made in the models described above about spin-orbit forces
in negative-parity excited baryons are likely to be premature, given
the information provided about convergence in Ref.~\cite {BB}.  It is
shown in the present work that the inclusion of negative-parity
excited baryons as intermediate states is crucial to the accurate
calculation of the mass shifts of these states.

It is, therefore, clear from considering prior work that a
self-consistent model of baryon self energies must employ a full set
of spin-flavor [SU(6)] symmetry related $B^\prime M$ intermediate
states, and at the same time must include excited baryon states up to
at least the $N=3$ band in order for the sum over intermediate state
baryons to have converged. This requires a detailed and universal
decay model, such as the $^3P_0$ pair-creation model, which is capable
of relating the baryon spectrum and the amplitudes for decay of a wide
variety of baryon initial states to a wide variety of baryon-meson
final states in an efficient way. The decay model needs to take into
account the spatial structure of the intermediate meson. It is also
clear that it will be necessary to modify the usual momentum
dependence of the decay amplitudes calculated in this model to take
into account the size of the constituent quark-pair creation vertex.

In addition, the size of these loop effects requires that the
interactions between the quarks required to fit the observed spectrum
be changed by their presence. To be consistent the wave functions used
to calculate the vertex amplitudes should then also be changed, and
the effect of these changes on the self-energies examined. The work of
Ref.~\cite{BB} has showed that the $\Delta$-nucleon splitting may not
be sensitive to such details, but from the sensitivity to the details
of the inter-quark Hamiltonian used to describe the hadron states
observed in many of these calculations, it can be expected that this
will be an important effect in the calculation of the self energies of
negative-parity excited baryons.
%
\section{\label{sec:Sigmas}Baryon self energies and bare energies}
%
In the present work, a calculation of the self energies of ground and
negative-parity excited $N$ and $\Delta$ states is carried out using a
$^3P_0$ pair-creation model to calculate the momentum-dependent
vertices ${\cal M}_{BB^\prime M}(k)$, with wave functions
calculated using a relativized model~\cite{CI} with a
variable-strength spin-dependent (one-gluon exchange) contact
interaction between the quarks. This calculation takes into account a
full set of spin-flavor symmetry related intermediate states
B$^\prime$M with
\begin{eqnarray}
\label{states}
{\rm M}&\in& \{\pi,K,\eta,\eta^\prime,\rho,\omega,K^*\} \nonumber\\
{\rm B^\prime}&\in& \{N,\Delta,N^*,\Delta^*,\Lambda,\Sigma,\Lambda^*,\Sigma^*\},
\end{eqnarray}
including all excitations of all of the intermediate baryon states up
to and including $N=3$ band states. Note that $\phi$ mesons couple
weakly to non-strange baryon states since such decays are OZI suppressed,
and so they are ignored.

The usual version of the $^3P_0$ model gives vertices that are too
hard, and the loop integrals required to evaluate the self energies
get large contributions from high momenta. Here these vertices are
modified by adopting a pair-creation operator used in previous
calculations of loop effects in mesons~\cite{GI} and
baryons~\cite{Silvestre-Brac:pw}. This operator includes a form factor
$\exp (-f^2[{\bf p}_q-{\bf p}_{\bar{q}}]^2)$ with $f^2=2.8$ GeV$^{-2}$,
which gives the quark-pair-creation vertex a size of around 0.33 fm.
As the self energies due to a given intermediate state depend
crucially on the masses adopted for the intermediate hadrons, these
are taken to be the physical masses, where known, and model
masses~\cite{CI} otherwise.

Since the self energies $\Sigma_{\rm B}(E)$ calculated using
Eq.~(\ref{SigmaB(E)}) are energy-dependent, it is necessary to solve
for the `bare' mass $E^0_B$ required to reproduce the known physical
masses $m_B$ of a given baryon $B$ by solving self-consistent
equations
\begin{equation}
E_B+\Sigma_B(E_B)=m_B.
\label{self-cons}
\end{equation}
This requires knowledge of the self energies at a range of bare
energies. For example, to examine the $\Delta-N$ splitting, the `bare'
masses required to reproduce the known physical masses of the $N$ and
$\Delta$ are found by solving a pair of (uncoupled) self-consistent
equations
\begin{equation}
E_N+\Sigma_N(E_N)=m_N,\ \ E_\Delta+\Sigma_{\Delta}(E_\Delta)=m_\Delta
\label{self-consND}
\end{equation}
for the `bare' masses $E^0_N$ and $E^0_\Delta$. Note that the self
energies tend to be large and negative, but only differences in the
self energies are observable.

Details of how the calculation is made are given in \Sec{sec:Method}.
Results for the self energies and resulting bare energies of ground
state and negative-parity excited state non-strange baryons are given
in \Secs{sec:ResultsD-N},~\ref{sec:ResultsP-waves}
and~\ref{sec:ResultsSpectra}. The conclusions of this study are given
in \Sec{sec:Concl}.
%
\section{\label{sec:Method}Baryon-meson intermediate state contributions}
%
In this section some of the formalism necessary to describe the
effects of baryon-meson loops on baryon masses is presented.
%
\subsection{\label{sec:vertex}Strong decay vertices}
%
A key ingredient of this calculation is the form of the momentum
dependence of the baryon-baryon-meson vertices. Here the $^3P_0$
pair-creation strong decay model is used to obtain the structure of
each vertex and hence its momentum dependence. The modified
pair-creation operator has the form
\begin{eqnarray} \label{3P0op}
T&=&-3\gamma\sum_{i,j}\int d {\bf p}_i d {\bf p}_j \ \delta( {\bf
p}_i + {\bf p}_j)\ C_{ij} \ F_{ij}\ e^{-f^2({\bf p}_i - {\bf
p}_j)^2}\nonumber\\ && \hphantom{\sum} \times \sum_m \langle 1,m;1,-m|0,0
\rangle \ \nonumber\\ && \hphantom{\sum \sum} \times \chi_{ij}^m \ { \cal
Y}_1^{-m}( {\bf p}_i - {\bf p}_j ) \ b_i^{\dagger}( {\bf p}_i ) \
d_j^{\dagger}( {\bf p}_j ),
\end{eqnarray}
where $C_{ij}$ and $F_{ij}$ are the color and flavor wave functions of
the created pair, both assumed to be singlet, $\chi_{ij}$ is the spin
triplet wave function of the pair, and ${\cal Y}_1({\bf p}_i-{\bf
p}_j)$ is the solid harmonic indicating that the pair is in a relative
$P$-wave $(l=1)$. Note that the threshold behavior resulting from the
$|{\bf p}_i-{\bf p}_j|$ factor in the solid harmonic is as seen
experimentally. Here $b_i^{\dagger}( {\bf p}_i\/ )$ and
$d_j^{\dagger}( {\bf p}_j\/)$ are the creation operators for a quark
and an antiquark with momenta ${\bf p}_i$ and ${\bf p}_j$,
respectively. As mentioned above, the additional exponential has been
introduced to give the vertex a spatial extent by creating the
quark-antiquark pair over a smeared region, instead of at a point as
is the case in the usual version of the model.

Baryon wave functions which result from diagonalizing the model $qqq$
Hamiltonian described in Sec.~\ref{sec:qqq-model} in a large harmonic
oscillator basis (up to and including the $N=7$ oscillator band) are
used along with single-oscillator meson wave functions to evaluate the
transition matrix elements of the pair-creation operator in
Eq.~(\ref{3P0op}). There are only two phenomenological parameters in
this model. These are $\gamma$, the $^3P_0$ coupling strength, which
is fitted to the experimentally well known $\Delta \rightarrow N \pi$
decay, and $f$, which is set to give a reasonable quark-pair-creation
vertex size. A similar model using the same wave functions
but~\cite{whyf=0?} with $f=0$ has been tested against a large number
of measured baryon decays~\cite{CR}. 

For the transition $B\rightarrow B^\prime M$, we are interested in
evaluating the transition amplitude
\begin{equation}\label{amp1}
A_{B \to B^\prime M}= \langle B^\prime M|T|B \rangle,
\end{equation}
which is given in Appendix~\ref{app:TA}. The notation illustrated in
Fig.~\ref{decay} was used to arrive at this form. Note that the
decaying baryon is assumed to be at rest and that the relative
momentum of the final baryon and meson is ${\bf k}_0$.

Given the very large number of loops which can contribute to the
self-energy of a given baryon~\cite{numloops}, and the requirement of
a calculation of the momentum dependent vertex ${\cal M}_{BB^\prime
M}(k)$ for each of them, this is a necessarily computationally
intensive calculation. Code has been written in {\it Maple}, and
executed on a computer cluster, which analytically calculates the
matrix elements of the operator in Eq.~(\ref{3P0op}) for each pair of
oscillator substates involved in the external and intermediate baryon
wave functions. This has the advantage that the momentum-dependent
vertices can then be repeatedly projected out of these stored matrices
with the external and intermediate baryon wave functions, which change
as the model $qqq$ Hamiltonian is adjusted. This process is described
in what follows.
%
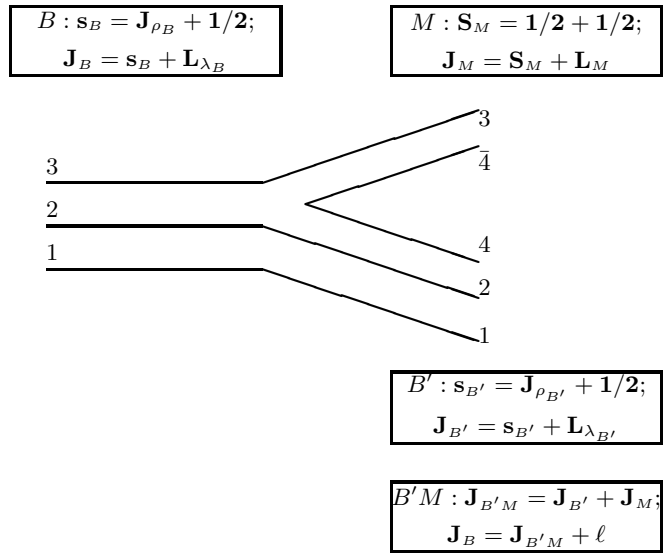
\begin{figure}
\hspace{-0.3in}
\input{decay_fig}
\caption{\label{decay}Schematic diagram of the decay $B\rightarrow
B'M$ in the $^3P_0$ model. The angular momentum notation is shown. The
decay proceeds through $B(123)\rightarrow 12(4\bar{4})3\rightarrow
B'(124)M(\bar{4}3)$.}
\end{figure}
%
\subsection{\label{sec:Self-cons}Self-consistent baryon mass calculation}
%
Given the expected size of splittings arising from self energies, it
will be necessary to adjust the $qqq$ Hamiltonian to fit the spectrum
of bare energies which result from fitting to baryon masses extracted
from analyses of scattering data. Each variation of the Hamiltonian
will produce a new set of baryon wave functions, which can in turn
be used to calculate the momentum dependence of the $BB^\prime M$
vertices. These will yield new self energies, from which a
new spectrum of bare masses required to fit the baryon masses from
analyses of data can be obtained [{\it i.e.} solutions of
Eq.~(\ref{self-cons})]. Obviously an iterative procedure will be
required to find the form of the $qqq$ Hamiltonian which best fits the
physical masses.

Given the complexity of the self-energy calculations, full
implementation of this iterative procedure is postponed until the set
of external states is expanded to include strange
baryons. Nevertheless, important conclusions about the strength of the
contact part of the spin-spin interaction will be made below by
comparison of the spectrum of a model $qqq$ Hamiltonian and the
consistently calculated bare energies. In addition, the effects of
configuration mixing due to a tensor interaction are explored in
Section~\ref{sec:ResultsP-waves}. In what follows, the form of the
model $qqq$ Hamiltonian used in the present calculations is described.
%
\subsection{\label{sec:qqq-model}Model $qqq$ Hamiltonian}
%
In order to calculate the strength of the baryon-baryon-meson
(BB$^\prime$M) vertices used here we require a model of the spectrum
and wave functions of the external (B) and intermediate-state
(B$^\prime$) baryons. While it may be possible to use physical masses
for those states present in analyses of scattering data, a model is
required to estimate the masses of states missing in these analyses
but present when baryons are composed of three quarks treated
symmetrically. 

It is also inconsistent to use physical masses for the intermediate
states along with harmonic-oscillator wave functions. The spectrum is
closer to that of a linear potential, and the wave functions will
include significant mixing due to this if expressed in a
harmonic-oscillator basis. After taking confinement into account by
means of a linear potential, most models have some sort of short-range
residual interaction between the quarks. It is interesting to explore
whether the self-energies of baryons depend on the presence of such
residual interactions.

In order to provide explicit wave functions roughly consistent with the
spectrum of the intermediate states, a model $qqq$ Hamiltonian with the
relativistic form of the kinetic energy and a linear confinement
potential
\begin{equation}
\label{H_0}
H_0=\sum_i\sqrt{p_i^2+m_i^2}+F\sum_{i<j}br_{ij}
\end{equation}
is used here, where $b=0.15$ GeV$^2$ is the baryon string tension and
$F=0.55$ is chosen to minimize the difference between this and the sum
of the lengths of a Y-shaped string~\cite{lattice-b} in a spherical
ground state (for details see Ref.~\cite{CI}). Note $r_{ij}=\vert {\bf
r}_i-{\bf r}_j\vert$ is the distance between quarks $i$ and $j$.

In order to explore the sensitivity of the self energies (or the bare
energies required to fit the physical masses extracted from analyses
of scattering data) to the presence of residual interactions, $H_0$ is
supplemented by pairwise Coulomb and contact interactions of a form
motivated by one-gluon exchange
\begin{equation}
H_{\rm Coulomb}+H_{\rm contact}=\sum_{i<j}H_{ij},
\end{equation}
where 
\begin{eqnarray}
\label{Hcont}
\nonumber
&H_{ij}=\left(\beta_{ij}\right)^{1/2+\epsilon_{\rm Coul}}
{\tilde G}(r_{ij})
\left(\beta_{ij}\right)^{1/2+\epsilon_{\rm Coul}}\\
&+\left(\delta_{ij}\right)^{1/2+\epsilon_{\rm cont}}
\left[{2{\bf S}_i\cdot{\bf S}_j\over 3m_im_j}
\nabla^2{\tilde G}(r_{ij})\right]
\left(\delta_{ij}\right)^{1/2+\epsilon_{\rm cont}}.
\end{eqnarray}
The outer factors of powers of
\begin{equation}
\delta_{ij}
=m_im_j/\left(\sqrt{p_{ij}^2+m_i^2}\sqrt{p_{ij}^2+m_j^2}\right)
\end{equation}
and
\begin{equation}
\beta_{ij}
=1+p_{ij}^2/\left(\sqrt{p_{ij}^2+m_i^2}\sqrt{p_{ij}^2+m_j^2}\right),
\end{equation}
where $p_{ij}$ is the magnitude of the momentum of the interacting
quarks in their center-of-momentum frame, are momentum-dependent
relativistic factors designed to parameterize the dependence of these
potentials away from the non relativistic limit, and are based on the
momentum dependence of the one-gluon exchange $T$-matrix element
between free quarks, where $\epsilon_{\rm cont}=\epsilon_{\rm
Coul}=0$. For bound quarks this momentum dependence will in general be
modified. In the following, $\epsilon_{\rm cont}=-0.168$ as in
Ref.~\cite{CI}, and $\epsilon_{\rm Coul}$ is given the same value for
simplicity.

In Eq.~(\ref{Hcont})
\begin{equation}
\label{Gtilde}
{\tilde G}(r_{ij})=-\sum_k {2\alpha_k\over 3r_{ij}}
{\rm erf}(\sigma_{kij}r_{ij})
\end{equation}
is a Coulomb potential smeared over a Gaussian distribution 
\begin{equation}
\rho_{ij}({\bf r}_{ij})
={\sigma_{ij}^2\over \pi^{3/2}}e^{-\sigma_{ij}^2
r_{ij}^2}
\end{equation}
of the inter-quark coordinate, with smearing parameters $\sigma_{ij}$
which depend on the masses of the interacting quarks via (for details see
Ref.~\cite{CI})
\begin{equation}
\sigma_{ij}^2={\sigma_0^2\over 2}
\left\{1+\left[{4m_im_j\over (m_i+m_j)^2}\right]^4\right\}
+s^2\left[{2m_im_j\over m_i+m_j}\right]^2,
\end{equation}
where $\sigma_0$ and $s$ are universal parameters. 
Note also that the strong coupling runs according to the usual perturbative
formula at large $Q^2$, and saturates to a value $\alpha_s^{\rm
critical}$ at low $Q^2$. This behavior is fit to a functional form
\begin{equation}
\alpha_s(Q^2)=\sum_{k=1}^3\alpha_k e^{-Q^2/4\gamma_k^2}
\end{equation}
and the parameters $\sigma_{kij}$ in Eq.~(\ref{Gtilde}) have values given by
\begin{equation}
1/\sigma_{kij}^2=1/\gamma_k^2+1/\sigma_{ij}^2.
\end{equation}

In Ref.~\cite{CI} the smearing parameters relevant to this work were
$\sigma_{uu}=\sigma_{ud}=\sigma_{dd}=1.832$ GeV, and
$\sigma_{us}=\sigma_{ds}=1.702$ GeV, resulting from $\sigma_0=1.8$ GeV
and $s=1.55$. This results in an effective `size' of a constituent
quark of roughly 0.08 fm. In related work it has been shown that, in
order to fit the nucleon electromagnetic form factors in a light-cone
based model using the wave functions which result from the Hamiltonian
used in Ref.~\cite{CI}, form factors for the constituent quarks are
required which give them an electromagnetic size much larger than
this~\cite{CPSS}. This is because the contact interaction of
Eq.~(\ref{Hcont}), which is a smeared Dirac $\delta$ function, has most
of its strength at short range, and so builds substantial high-momentum
components into the wave functions of states like the nucleon with net
attractive contact interactions. A constituent quark form factor which
falls off rapidly with momentum transfer is required to compensate. A
simple solution to this apparent mismatch of the strong and
electromagnetic sizes of the constituent quark, which is adopted here,
is to use a parameter $\sigma_0=0.83$ GeV, which gives the up and down
quark interactions a smearing $\sigma=0.9$ GeV, or a constituent quark
size of roughly 0.15 fm.

Accompanying this increase in the effective range of the contact
interaction is a reduction in its average strength, which can be
quantified by the splitting caused by $H_{\rm contact}$ between the
ground-state $\Delta$ and nucleon. As a consequence, if the value
$\alpha_s^{\rm critical}=0.60$ from Ref.~\cite{CI} is used with this
longer-range contact interaction, the size of the $\Delta-N$ splitting
due to $H_{\rm contact}$ is reduced by about a factor of two. In what
follows a value $\alpha_s^{\rm critical}=0.55$ was used whenever
$H_{\rm Coulomb}+H_{\rm contact}$ was included in the model $qqq$
Hamiltonian.

Tensor interactions between the quarks are also considered, in order
to examine the effects of configuration mixing in the wave functions of
spin-partner states such as $N1/2^-(1535)$ and $N1/2^-(1650)$, which
have quark spin $1/2$ and $3/2$ respectively when the interactions
between quarks are overall spin scalars (such as $H_{\rm
contact}$). These can be consistently included by adding an
interaction
\begin{widetext}
\begin{equation}
H_{\rm tensor}=
\sum_{i<j}
\left(\delta_{ij}\right)^{1/2+\epsilon_{\rm tens}}
{1\over 3m_im_j}
\left[{3{\bf S}_i\cdot{\bf r}_{ij}{\bf S}_j\cdot{\bf r}_{ij}\over r_{ij}^2}
-{\bf S}_i\cdot{\bf S}_j\right]
\left[
{1\over r_{ij}}{d{\tilde G}(r_{ij})\over dr_{ij}}-
{d^2{\tilde G}(r_{ij})\over dr_{ij}^2}
\right]
\left(\delta_{ij}\right)^{1/2+\epsilon_{\rm tens}},
\end{equation}
\end{widetext}
and in what follows we have chosen $\epsilon_{\rm tens}$ to be the
same as $\epsilon_{\rm cont}$ and $\epsilon_{\rm Coul}$ for
simplicity.
%
\section{\label{sec:ResultsD-N}$\Delta$-Nucleon splitting}
%
The $\Delta$-Nucleon splitting has been used by those constructing
models of the baryon spectrum to determine the strength of the
short-range interactions between the quarks. It is therefore of
considerable interest to examine whether this splitting is modified by
the self energies which result from the presence of $B^\prime M$
intermediate states. Prior calculations show a substantial splitting
between the bare energies required to fit the physical $\Delta$ and
nucleon masses~\cite{PZ,BB,BHM}, but these calculations may not have
converged due to the restriction of the intermediate states to ground
state baryons~\cite{PZ,BHM}, or to non-strange baryons and
pions~\cite{BB}. Here this splitting is re-examined without these
restrictions, using wave functions generated by the Hamiltonian $H_0$
of Eq.~(\ref{H_0}) without residual interactions between the quarks,
and also those found using the reduced strength one-gluon exchange
interaction (with and without consistent tensor interactions and the
configuration mixing they cause) described in
Sec.~\ref{sec:qqq-model}. It is of particular interest to see whether
the sum over intermediate states in Eq.~(\ref{SigmaB(E)}) has
converged to a stable $\Delta$-Nucleon bare mass splitting, and
whether this splitting depends on the nature of these residual
interactions, as seen in Ref.~\cite{BB}.

\begin{figure}
\includegraphics[width=6.9cm,angle=-90]{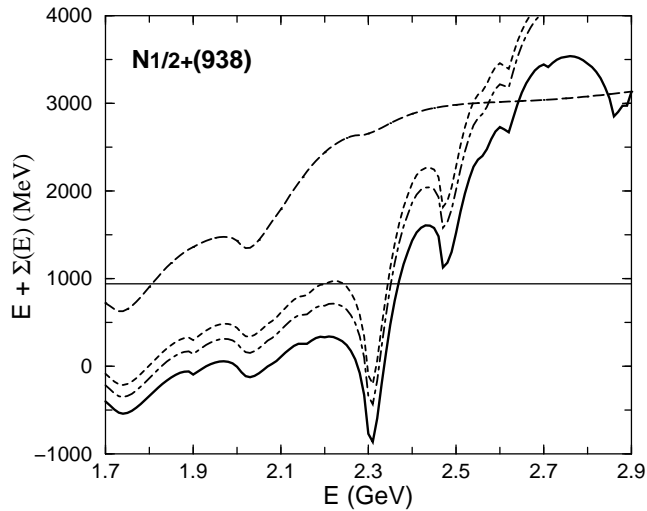}
\caption{\label{fig:n1p1} Sum of bare and self energies for the
ground-state nucleon, for $\alpha=0.5$ GeV and with Coulomb and
contact interactions only in $H_{qqq}$. The long-dashed curve is
calculated with only ground state intermediate baryons B$^\prime$, the
short-dashed curve adds $N=1$ band baryons, the short-dashed
long-dashed curve adds $N=2$ baryons, and the solid curve adds $N=3$
intermediate baryons. The first of Eqs.~(\protect{\ref{self-consND}})
is solved where the curves intersect the horizontal solid line at
$m_N=938$ MeV.}
\end{figure}

\begin{figure}
\includegraphics[width=6.9cm,angle=-90]{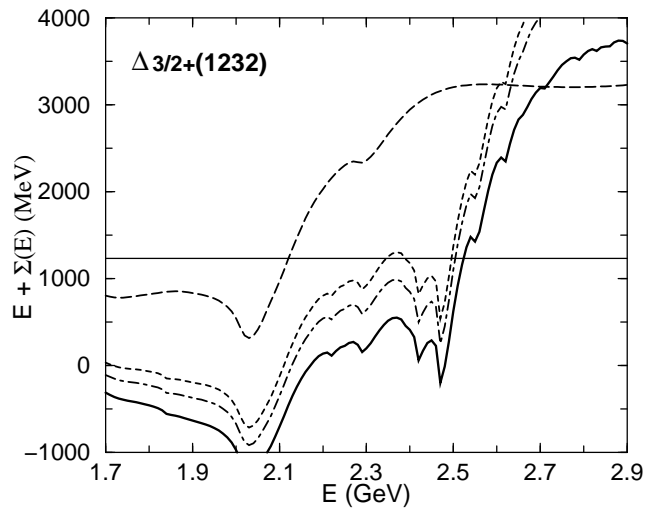}
\caption{\label{fig:d3p1} Sum of bare and self energies for the
ground-state $\Delta$, for $\alpha=0.5$ GeV and with Coulomb and contact
interactions only in $H_{qqq}$. Curves are labeled as in
Fig.~\protect{\ref{fig:n1p1}}.  The second of
Eqs.~(\protect{\ref{self-consND}}) is solved where the curves intersect
the horizontal solid line at $m_\Delta=1232$ MeV.}
\end{figure}

When baryon wave functions resulting from the $qqq$ Hamiltonian $H_0$
(only confining interactions between the quarks) are used to describe
the full set of $B^\prime M$ intermediate states in
Eq.~(\ref{states}), including excited baryon states up to the $N=3$
band, this results in bare masses which satisfy
$E^0_\Delta-E^0_N\simeq 150$ MeV. The addition of the reduced strength
Coulomb and contact interactions described in
Section~\ref{sec:qqq-model} changes this slightly to roughly 155
MeV. The results described below use the latter Hamiltonian except
where noted. Figures~\ref{fig:n1p1} and~\ref{fig:d3p1} illustrate the
energy dependence of the self energy of the nucleon and $\Delta$
ground states, by plotting $E+\Sigma(E)$ against $E$.
Eqs.~(\ref{self-consND}) are solved where the curves intersect the
horizontal solid lines at $m_N=938$ MeV and $m_\Delta=1232$ MeV. The
four curves show the effects of increasing the maximum level of
excitation of the intermediate baryons from ground states ($N_{\rm
max}=0$), to $N_{\rm max}=3$, with regions of rapid curvature
corresponding to various decay thresholds.

The first row of Table~\ref{splittings} illustrates the dependence of
the difference in the bare energies $E^0_\Delta-E^0_N$ which solve
Equations~(\ref{self-consND}) on the maximum level of excitation of the
intermediate baryon states. This and Figs.~\ref{fig:n1p1}
and~\ref{fig:d3p1} show the importance of the inclusion of
intermediate states involving ``$N=1$ band'' negative-parity excited
baryons, i.e. those which have wave functions predominantly made up of
$N=1$ band oscillator substates. These results confirm those of
Ref.~\cite{BB}, where it is shown, in a model with the intermediate
states restricted to pions and ground and excited states of $N$ and
$\Delta$, that the inclusion of excited baryon intermediate states
substantially reduces the $\Delta-N$ splitting.

In the present work intermediate states involving $N=2$ and $N=3$ band
excited baryon states are relatively unimportant, contributing less
than 10 MeV to the splitting. These results clearly demonstrate that,
by $N_{\rm max}=3$, the difference of the self energies of the
$\Delta$ and nucleon has converged, as seen in the quite different
model of these effects in Ref.~\cite{BB}. Note that this convergence
occurs even with the large increase in the multiplicity of intermediate baryon
states at higher $N$ values.
%
\begin{table}
\caption{\label{splittings}Representative splittings in MeV of bare energies
as a function of the maximum level $N_{\rm max}$ of excitation of the
intermediate baryons $B^\prime$, calculated using $\alpha=0.5$ GeV and
with Coulomb and contact interactions only in $H_{qqq}$.}
\begin{ruledtabular}
\begin{tabular}{lcccc}
 & $N_{\rm max}=0$ & $N_{\rm max}=1$ &$N_{\rm max}=2$ & $N_{\rm max}=3$\\
\hline
$E^0_\Delta-E^0_N$ & 315 & 152 & 153 & 155 \\
$E^0_{1650}-E^0_{1535}$ & 33 & 363 & -25 & -25 \\
$E^0_{1700}-E^0_{1520}$ & 25 & 280 & -17 & -25 \\
$P$-wave$-$grd. state & -119 & -49 & 339 & 334 \\
\end{tabular}
\end{ruledtabular}
\end{table}
%

Table~\ref{D-N} shows the difference of the self energies of the
$\Delta$ and nucleon, $\Sigma_\Delta(E^0_\Delta)-\Sigma_N(E^0_N)$,
broken up into contributions from intermediate states of different
flavor and level of baryon excitation. It is important to note that
all of these (energy-dependent) self-energy differences are evaluated
at the bare energies which solve Eqs.~(\ref{self-consND}) with the
complete sum of SU(6)-related intermediate states including baryon
excitations up to N=3. Their sum is therefore the difference of the
physical $\Delta-N\simeq 295$ MeV splitting and the $N_{\rm max}=3$
result from Table~\ref{splittings}. As these differences are strongly
energy dependent and the bare energies depend on the maximum level of
excitation of the intermediate-state baryons, the convergence of the
$\Delta-N$ splitting is demonstrated by reading Table~\ref{splittings}
from left to right, {\it not} Table~\ref{D-N}.

What is clear from Table~\ref{D-N} is that intermediate $B^\prime \pi$
states (with $B^\prime$ taken from ground and excited states of the
nucleon and $\Delta$) will contribute a self-energy difference only a
little larger than the full result. Iintermediate states involving
other pseudoscalar mesons add another 30 MeV, anthe additional terms
due to intermediate states involving all of the vector mesons $\rho$,
$\omega$ and $K^*$ reduce the sum by 50 MeV. Intermediate states
involving ground and excited state baryons and vector mesons are
clearly important. Interestingly, although the self-energy difference
due to sets of intermediate states involving $\rho$, $\omega$ and
$K^*$ mesons are individually large, especially when accompanied by
$N=1$ band baryons, their sum is not. This is reminiscent of results
for $\rho-\omega$ splitting in the very similar model of
Ref.~\cite{GI}, where it was shown that there are meson-meson
intermediate states which give large contributions to the splitting,
but which largely cancel when considered in certain groups.
%
\begin{table}
\caption{\label{D-N}Contributions to the difference
$\Sigma_\Delta(E^0_\Delta)-\Sigma_N(E^0_N)$ in MeV of the self
energies of the $\Delta$ and nucleon, evaluated at the full bare energies
$E^0_\Delta$ and $E^0_N$. Self energies are calculated using $\alpha=0.5$ GeV
and with Coulomb and contact interactions only in $H_{qqq}$. Columns
correspond to the excitation level of the intermediate baryons, and
the row labeled $\pi$ includes
contributions from $N^*\pi$ and $\Delta^*\pi$ intermediate states,
that labeled $K$ includes contributions from $\Lambda^* K$ and
$\Sigma^* K$ intermediate states, etc.}
\begin{ruledtabular}
\begin{tabular}{lcccc|c}
 &$N=0$ & $N=1$ &$N=2$ & $N=3$ & total\\
\hline
$\pi$         & -24 &   55 &    7 &  113 &  151\\
$K$           &  -4 &   27 &   12 &  -25 &   10\\
$\eta$        &   9 &  -14 &   -4 &  -13 &  -22\\
$\eta^\prime$ &  22 &   26 &   -3 &   -3 &   42\\
$\rho$        & -17 &  333 &  -16 &  -76 &  224\\
$\omega$      &  87 &  130 &  -10 &  -23 &  184\\
$K^*$         & 175 & -569 &  -34 &  -24 & -452\\
\hline
total         & 248 &  -12 &  -48 &  -51 &  137\\
\end{tabular}
\end{ruledtabular}
\end{table}
%

%
Similar conclusions resulted from a prior calculation which did not
use a full set of SU(6)-related $B^\prime M$ intermediate states,
where substantial $E^0_\Delta-E^0_N$ splittings were found using
wave functions including reduced-strength one-gluon exchange
interactions~\cite{BB}. In addition, this agrees reasonably well with
the expectation from a model with both one-pion exchange and one-gluon
exchange residual interactions between the
quarks~\cite{Buchmann:1996bd}, that about two thirds of the
$\Delta$-$N$ splitting comes from one-gluon exchange
interactions. However, in the present picture the rest of the
splitting arises from a source very different from one-boson-exchange
(OBE) or similar mechanisms between the quarks. Note that the
self-energy diagrams evaluated in this work include meson-exchange
interactions between the quarks, and also deal consistently with quark
self energies from meson loops and the threshold effects of a large
number of $B^\prime M$ intermediate states.

Table~\ref{D-Ncont} illustrates that the introduction of residual
interactions which can accommodate the rest of the observed
$\Delta$-$N$ splitting affects the wave functions and so the vertices,
but does not significantly affect the difference in the bare energies
of the ground state $N$ and $\Delta$. The latter is in contrast to the
results of Ref.~\cite{BB}, and this difference may be due to the
restricted set of intermediate states in that calculation, and also
the use of model masses for the intermediate states, which moves
thresholds away from their physical positions. It will be shown in the
next section that in the present calculation the differences of the
self energies of the negative-parity excited states do depend on the
residual interactions between the quarks.

%
\begin{table}
\caption{\label{D-Ncont}Splitting $E^0_\Delta-E^0_N$ in MeV of the
bare $\Delta$ and nucleon energies calculated using wave functions with
$H_{qqq}$ equal to $H_0$ (with $\alpha=0.4$ GeV), $H_0+H_{\rm
Coulomb}+H_{\rm contact}$, and $H_0+H_{\rm Coulomb}+H_{\rm
contact}+H_{\rm tensor}$ (with $\alpha=0.5$ GeV), and a full set of
intermediate states with excited baryons up to $N_{\rm max}=3$.}
\begin{ruledtabular}
\begin{tabular}{l|ccc}
$H_{qqq}$ & $H_0$ & $H_0+H_{\rm Coulomb}$ & $H_0+H_{\rm Coulomb}$ \\
 & & $+H_{\rm contact}$ & $+H_{\rm contact}+H_{\rm tensor}$ \\
\hline
$E^0_\Delta-E^0_N$ & 151 & 155 & 145 \\
\end{tabular}
\end{ruledtabular}
\end{table}
%
%
\section{\label{sec:ResultsP-waves}Non-strange $P$-wave baryon splittings}
\begin{figure}
\includegraphics[width=7.6cm,angle=-90]{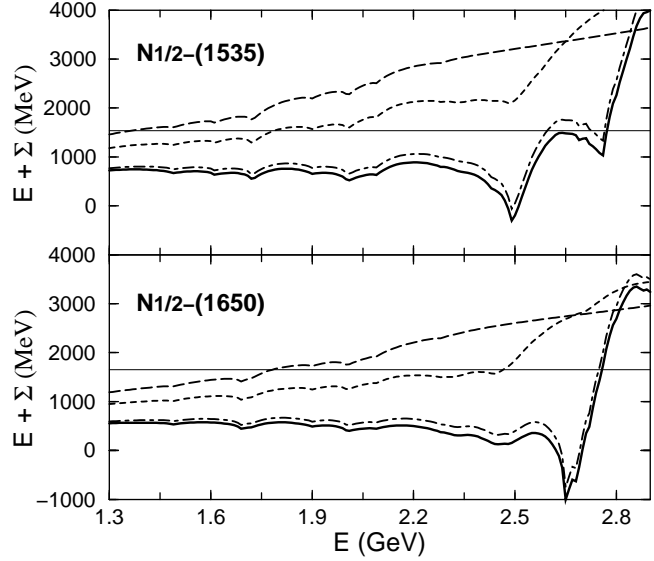}
\caption{\label{fig:n1m1_n1m2} Sum of bare and self energies for the
$P$-wave excited states $N1/2^-(1535)$ (upper panel) and
$N1/2^-(1650)$ (lower panel), for $\alpha=0.5$ GeV and with Coulomb
and contact interactions only in $H_{qqq}$. Curves are labeled as in
Fig.~\protect{\ref{fig:n1p1}}. Note Eq.~(\protect{\ref{self-cons}}) is
solved where the curves intersect the horizontal solid lines at 1535
and 1650 MeV, respectively.}
\end{figure}

%
The solution of Eq.~(\ref{self-cons}) for the bare energies of the
lowest-lying negative-parity $N^*$ states with $J^P=1/2^-$,
corresponding to the states $N(1535)S_{11}$ and $N(1650)S_{11}$ seen
in analyses of pion-nucleon scattering and photo-production data, is
illustrated in Figure~\ref{fig:n1m1_n1m2}. It is clear that the
inclusion of intermediate states involving both $N=1$ negative-parity
and $N=2$ band positive-parity excited baryon states is crucial to the
correct description of the bare masses of these states. Given the size
of the self energies due to intermediate states involving $N=2$ band
baryons, it was necessary to calculate the effects on the bare masses
of the presence of intermediate states involving $N=3$ band
highly-excited negative-parity baryon states. It is clear from
Fig.~\ref{fig:n1m1_n1m2} and Table~\ref{splittings} that these effects
are roughly the same for both states, which means that their splitting
is not strongly affected. Table~\ref{splittings} shows that, with
Coulomb and contact interactions only in the $qqq$ Hamiltonian which
generates the wave functions used to evaluate the self energies, the
bare mass of the $S_{11}(1650)$ state lies slightly below that of
$S_{11}(1535)$. A similar situation arises for the pair of states
$D_{13}(1520)$ and $D_{13}(1700)$.
%
%
When this process is repeated for the other negative-parity non strange
excited states, it is found that the bare energies required to fit the
physical masses are not degenerate. 

\begin{figure}
\includegraphics[width=8.6cm]{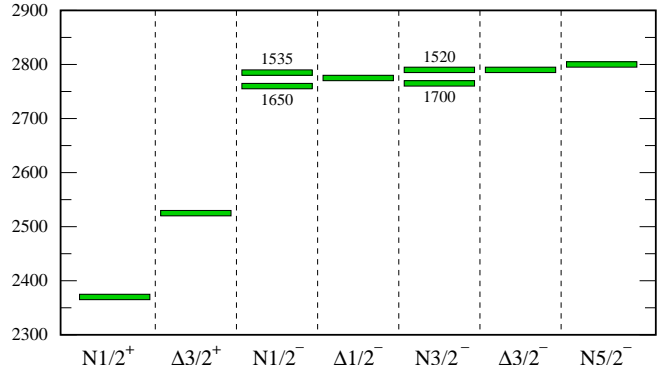}
\caption{\label{fig:spectrum-c} Bare energies (in MeV) of ground and
$P$-wave excited state non-strange baryons required to fit their
masses from analyses of data, calculated using wave functions which
are eigenfunctions of $H_0+H_{\rm Coulomb}+H_{\rm contact}$, with
harmonic-oscillator size parameter $\alpha=0.5$.}
\end{figure}

The pattern of splitting of the bare energies of these states and
those of the nucleon and $\Delta$ ground states is shown in
Figure~\ref{fig:spectrum-c}. Interestingly, although inverted from the
usual OGE quark-model expectations (where the predominantly spin-3/2
state lies above the predominantly spin-1/2 state), the bare mass
splitting required to fit the physical masses of the two $N1/2^-$
states $N(1535)S_{11}$ and $N(1650)S_{11}$ is considerably smaller
(about one third) than the physical mass splitting, similar to what
was found in the $\Delta$-nucleon ground state system. Put another
way, this means that the corrections to the mass from the self
energies resemble the splittings which arise from one-gluon exchange
or other spin-dependent contact interactions. Note also that there are
some effects which resemble tensor or spin-orbit interactions, such as
the small bare mass splitting between the $\Delta 1/2^-$ and $\Delta
3/2^-$ states.
%
%

It is interesting to determine whether the bare mass spectrum required
to fit the masses from analyses of scattering data depends on the
presence of configuration mixing effects in the wave functions used to
determine the $BB^\prime M$ vertices and so the self energies. A
simple way to test this, adopted here, is to include
one-gluon-exchange (OGE) tensor interactions in the $qqq$ Hamiltonian
used to calculate the wave functions (and model masses) of the
intermediate baryon states. In anticipation of the results of this
calculation, one could argue that changes in the wave functions of the
ground states from tensor interactions are at the level of a few
percent. These changes are unlikely to have large effects on the bare
masses of the nucleon and $\Delta$ ground states, because most of
their self energies arises from intermediate states which include
ground-state and low-lying negative-parity baryons. 

The self energies due to ground state intermediate baryons will likely
be unaffected.  Although it is known~\cite{IKnegpar} that large
mixings of the negative-parity states arise from tensor interactions,
the self energies of the nucleon and $\Delta$ ground states due to
these intermediate states should also largely be unaffected, because
these states are close to degenerate on the scale of the mass
splitting between the ground states and the negative-parity excited
states. Mixings, therefore, will shift strength around between
individual intermediate states, but in this case the energy
denominators in the loop integrals in Eq.~(\ref{SigmaB(E)}) are roughly
the same for each intermediate state.

\begin{figure}
\includegraphics[width=8.6cm]{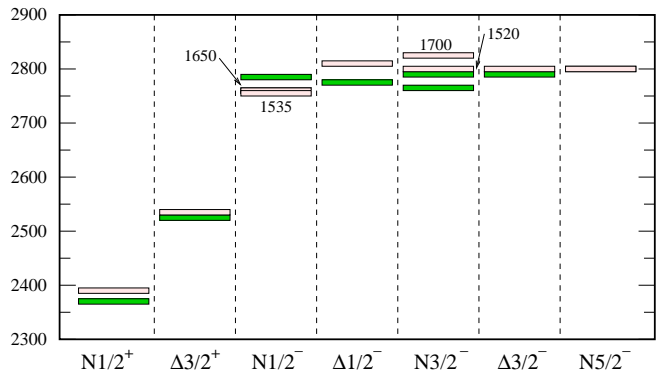}
\caption{\label{fig:spectrum-ct} Bare energies (in MeV) of ground and
$P$-wave excited state non-strange baryons required to fit their
masses from analyses of data, calculated using wave functions which
are eigenfunctions of $H_0+H_{\rm Coulomb}+H_{\rm contact}+H_{\rm
tensor}$ (lightly shaded boxes), compared to those calculated using
eigenfunctions of $H_0+H_{\rm Coulomb}+H_{\rm contact}$ (dark-shaded
boxes, see Fig.~\protect{\ref{fig:spectrum-c}}), with
harmonic-oscillator size parameter $\alpha=0.5$.}
\end{figure}

This will not be true of the self energies of negative-parity excited
states, where the mixings will have substantial effects on the vertex
functions, and the energy denominators for $B\to B^\prime M\to B$ can
differ due to the proximity in mass of the initial and intermediate
states. The effects on the bare masses of including these mixings are
shown in Figure~\ref{fig:spectrum-ct}, where the spectrum of bare
masses calculated with tensor mixings in the wave functions is
contrasted to that from Fig.~\ref{fig:spectrum-c}. The bare mass
splitting between the ground state $\Delta$ and nucleon is slightly
reduced, but there are substantial changes in the bare masses of the
negative-parity states, and in the splitting between the average bare
mass of the ground state and negative-parity excited states. The bare
masses of the two $N1/2^-$ states are now in the usual order, if
almost degenerate, and the same is true of the two $N3/2^-$
states. There is a small negative splitting which resembles a
spin-orbit splitting between the bare masses of the $\Delta
3/2^-(1700)$ and $\Delta 1/2^-(1620)$ states which, interestingly, has
the opposite sign to that expected in the OGE quark model of
Ref.~\cite{CI}.
%
%
\section{\label{sec:ResultsSpectra}Comparison of bare mass and model spectra}
%
Given the substantial size of the splittings induced by baryon self
energies demonstrated above, it can be argued that it should not be
possible to fit the spectrum by ignoring them. Instead, it will be
necessary to (i) postulate a $qqq$ Hamiltonian, (ii) use the resulting
wave functions to calculate the self energies which result from it,
(iii) use these to find the bare baryon energies corresponding to the
physical masses, and (iv) check to see whether the splittings between
these bare energies match those between the model masses resulting
from step (i). This process may need to be iterated.

\begin{figure}
\includegraphics[width=8.6cm]{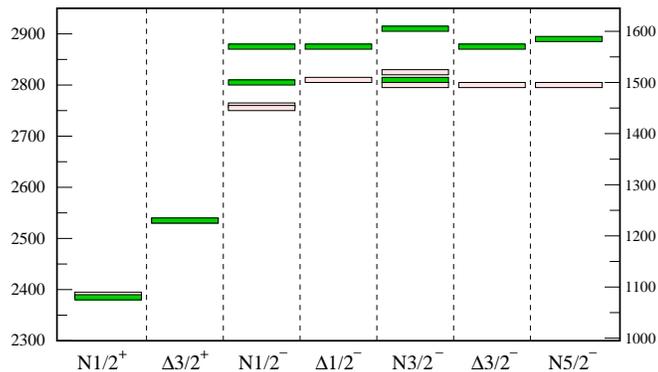}
\caption{\label{fig:model-bare} Bare energies (in MeV) of ground and
$P$-wave excited state non-strange baryons required to fit their
masses from analyses of data, calculated using wave functions which
are eigenfunctions of $H_0+H_{\rm contact}+H_{\rm tensor}$ (lightly
shaded boxes, left scale) with harmonic-oscillator size parameter
$\alpha=0.5$, compared to model masses (in MeV) from the same
Hamiltonian (dark-shaded boxes, right scale). An overall constant has
been added to the model masses to fit the $\Delta(1232)$ mass,
and the two scales have been adjusted so that the two $\Delta(1232)$
masses coincide.}
\end{figure}

However, prior calculations of the baryon spectrum which ignore the
self energies seem able to roughly fit the physical masses, with some
noticeable exceptions. As has been demonstrated above and in work by
other authors~\cite{PZ,BB,BHM,Silvestre-Brac:pw}, this may be because
the splittings in the bare masses often act in the same direction as
spin-dependent contact interactions between the
quarks. Figure~\ref{fig:model-bare} and Table~\ref{Spectrum} show the
comparison between the model masses resulting from the one-gluon
exchange Hamiltonian described in Sec.~\ref{sec:qqq-model} and the
corresponding bare masses. This comparison represents the second
iteration in this process, where the strength of the
one-gluon-exchange interaction has been reduced to roughly fit the
$\Delta-N$ bare-mass splitting calculated consistently with the
corresponding wave functions.

It is obvious from Fig.~\ref{fig:model-bare} that the string tension
(or light quark mass) used in Ref.~\cite{CI} and adopted here is not
able to fit the splitting in the average bare masses of the ground
states and negative-parity excited states. (This splitting does not
change significantly with the addition of intermediate states
involving $N=3$ band baryons, as illustrated in
Table~\ref{splittings}). It is possible to self-consistently fit the
splitting of the $\Delta$ and nucleon bare masses, which is a
non-trivial result, as pointed out in
Ref.~\cite{BB}. Figure~\ref{fig:model-bare} shows that the resulting
model $qqq$ Hamiltonian, if calculated with a consistent tensor
interaction, gives model splittings which resemble those of the
consistently calculated bare masses, with some differences. Although
beyond the scope of the present work, given the sensitivity of the
results for bare masses to the presence of the tensor interactions
demonstrated above, and the results of previous
calculations~\cite{BHM,Silvestre-Brac:pw,Fujiwara}, it will be
interesting to self-consistently calculate the effects of spin-orbit
interactions.
%
\begin{table}
\caption{\label{Spectrum}Spectrum in MeV of bare energies calculated
using wave functions with $H_{qqq}=H_0+H_{\rm Coulomb}+H_{\rm
contact}+H_{\rm tensor}$, and the corresponding model masses. Self
energies are calculated using a full set of intermediate states with
excited baryons up to $N_{\rm max}=3$. Both spectra have been
normalized to reproduce the mass of $\Delta(1232)$ by adjusting an
overall additive constant.}
\begin{ruledtabular}
\begin{tabular}{lcc}
physical state & $E^0$ & model mass\\
\hline
$N{1\over 2}^+(938)$ & 1087 & 1082 \\
$\Delta{3\over 2}^+(1232)$ & 1232 & 1232 \\
$N{1\over 2}^-(1535)$ & 1453 & 1500 \\
$N{1\over 2}^-(1650)$ & 1457 & 1572 \\
$\Delta{1\over 2}^-(1620)$ & 1507 & 1570 \\
$N{3\over 2}^-(1520)$ & 1495 & 1506 \\
$N{3\over 2}^-(1700)$ & 1520 & 1606 \\
$\Delta{3\over 2}^-(1700)$ & 1495 & 1569 \\
$N{5\over 2}^-(1675)$ & 1495 & 1584 \\
\end{tabular}
\end{ruledtabular}
\end{table}

%
\section{\label{sec:Concl}Conclusions}
%
The results shown above demonstrate that the sum over excited baryons
in the intermediate state has converged to a stable splitting between
the bare energies required to fit the physical ground-state nucleon
and $\Delta$ masses. This requires the use of a full set of
SU(6)-related $B^\prime M$ intermediate states, with excited baryons
$B^\prime$ up to the top of the $N=3$ band. This splitting is roughly
150 MeV, independent of the choice of inter-quark Hamiltonian
$H_{qqq}$ used to generate the wave functions, which affect the
$BB^\prime M$ vertices used to evaluate the self energies. In prior
calculations of the baryon spectrum with one-gluon-exchange residual
interactions between the quarks, this splitting has been used to fix
the effective strength $\alpha_s$ of the short-range interactions
between the quarks. This result implies that a reduced-strength
one-gluon-exchange interaction should be employed in such
calculations, designed to fit the roughly 145 MeV splitting between
these states after the self-consistent correction for the self
energies has been applied. Similar conclusions resulted from
calculations which either did not use a full set of SU(6)-related
$B^\prime M$ intermediate states or did not consider spatially excited
intermediate baryons. The self-energy diagrams evaluated in this work
include meson-exchange interactions between the quarks, and deal
consistently with quark self energies from meson loops and the
threshold effects of $B^\prime M$ states. With this complexity in
mind, it may still be possible to conclude that such meson-exchange
effects are not the sole source of the $\Delta-N$ splitting.

Convergence of the sum over intermediate excited baryons of the bare
energies required to fit the $P$-wave non strange baryon spectrum has
been demonstrated using this same set of $B^\prime M$ states, although
in contrast to the situation with the ground states, $N=2$ band states
make important contributions. This resolves a problem with convergence
found in a prior calculation, and points to the importance of taking
into account the structure of the intermediate-state mesons, and the
suppression of the creation of quark pairs with high relative
momentum, as adopted previously by other authors. Interestingly, these
bare energies are not degenerate, and also depend substantially on the
inter-quark Hamiltonian used to generate the wave functions which go
into the calculation of the $BB^\prime M$ vertices in the self
energies. This means that it is necessary to use in the evaluation of
these self energies a self-consistent calculation of both the
spectrum, with its corresponding wave functions, and the strong
vertices. The use in such calculations of unmixed oscillator
wave functions to represent states whose physical masses enter into the
positions of thresholds is inconsistent, and is likely to lead to
erroneous results.

These observations suggest the use of a reduced-strength one-gluon
exchange interaction to generate the wave functions used in the
calculation of the vertices.  The comparison of the resulting model
spectrum with the bare energies required to fit the spectrum of ground
and $P$-wave non strange baryons demonstrates that it is possible to
understand the $\Delta-N$ splitting as partly due to self-energies and
partly due to residual interactions between the quarks. Calculations
which ignore the self energies overestimate the splitting between the
ground states and the $P$-wave excited states. The comparison between
the splittings of the $P$-wave baryon model masses and those of the
resulting bare masses shows that tensor mixings are important, and
this is likely to also be true of spin-orbit mixings.

An extension of the present work to the ground-state and
negative-parity excited state $\Lambda$ and $\Sigma$ baryons along
with a study of spin-orbit effects is currently in progress. This is
of particular interest since intermediate states such as $\bar{K}N$
are known to have large effects on the masses and properties of
several of the excited states, such as $\Lambda1/2^-(1405)$. The
mixing of both non-strange and strange baryon states with the same
quantum numbers due to these $B^\prime M$ intermediate states, ignored
here for simplicity, is also under investigation, as are the effects
of self-energies on positive-parity excited initial baryon states
(such as the Roper resonance). This will likely require the inclusion
of a second band of positive-parity excited states. At the same time
the effects of excited meson states, presumed small for the reasons
outlined above, should be checked by examining self energies due to
the lightest orbitally excited mesons and ground-state baryons.
\begin{acknowledgments}
The authors wish to thank Winston Roberts for his invaluable help with
understanding the calculation of baryon strong decays, and Nathan
Isgur, Paul Geiger, and Eric Swanson for illuminating discussions.
This work is supported by the U.S. Department of Energy under contract
DE-FG02-86ER40273. The calculations required to complete this work
were carried out on a cluster of computers provided to the Department
of Physics by the College of Arts and Sciences of the Florida State
University.
\end{acknowledgments}

\appendix

%
\appendix
\section{\label{app:TA}Transition Amplitude}
All of the details of the calculation of the strong decay transition
amplitudes used in the present work are given
elsewhere~\cite{RS-B,DMthesis}, including the full form of the various
components of the decay amplitude~\cite{CR}. A summary
of the main results used in their evaluation is given below.

The final form of the transition amplitude is
\begin{widetext}
\begin{eqnarray} \label{ampa1}      
A_{B \to B^\prime M}&=&{6\gamma \over 3 \sqrt{3}}(-1)^{J_{\scriptscriptstyle
B}+J_{\scriptscriptstyle B^\prime}+\ell_{\scriptscriptstyle 
B}+\ell_{\scriptscriptstyle B^\prime}-1} \sum_{J_{\rho},s_{\scriptscriptstyle 
B},s_{\scriptscriptstyle B^\prime}}
 {\hat J}_{\rho}^2 {\hat s}_{\scriptscriptstyle B} {\hat 
S}_{\scriptscriptstyle B} {\hat L}_{\scriptscriptstyle B} {\hat
s}_{\scriptscriptstyle B^\prime} {\hat S}_{\scriptscriptstyle B^\prime} 
{\hat L}_{\scriptscriptstyle B^\prime}
\begin{Bmatrix}S_{\scriptscriptstyle
B}&L_\rho&s_{\scriptscriptstyle B} \\
               \ell_{\scriptscriptstyle B}&J_{\scriptscriptstyle
B}&L_{\scriptscriptstyle B} \end{Bmatrix}
\begin{Bmatrix}L_\rho&S_\rho&J_\rho \\
               {1 \over 2}&s_{\scriptscriptstyle
B}&S_{\scriptscriptstyle B} \end{Bmatrix} \nonumber \\
&&\times \begin{Bmatrix}S_{\scriptscriptstyle B^\prime}&L_\rho&
s_{\scriptscriptstyle
B^\prime} \\
               \ell_{\scriptscriptstyle B^\prime}&J_{\scriptscriptstyle 
B^\prime}&L_{\scriptscriptstyle B^\prime} \end{Bmatrix}
\begin{Bmatrix}L_\rho&S_\rho&J_\rho  \\
               {1 \over 2}&s_{\scriptscriptstyle
B^\prime}&S_{\scriptscriptstyle B^\prime} \end{Bmatrix} 
(-1)^{\ell+\ell_{\scriptscriptstyle B}+J_{\scriptscriptstyle 
M}-L_{\scriptscriptstyle M}-S_{\scriptscriptstyle M}}{\cal F}(BB^\prime M) 
{\cal R}(BB^\prime M)\nonumber \\
&&\times \sum_{S_{\scriptscriptstyle B^\prime M}}(-1)^{s_{\scriptscriptstyle
B}-S_{\scriptscriptstyle B^\prime M}} 
\begin{bmatrix}J_\rho&1/2&s_{\scriptscriptstyle B^\prime} \\
               1/2&1/2&S_{\scriptscriptstyle M} \\
               s_{\scriptscriptstyle B}&1&S_{\scriptscriptstyle B^\prime M} \end{bmatrix}
\sum_{L_{\scriptscriptstyle B^\prime M}} (-1)^{L_{\scriptscriptstyle B^\prime 
M}}
\begin{bmatrix}s_{\scriptscriptstyle B^\prime}&\ell_{\scriptscriptstyle 
B^\prime}&J_{\scriptscriptstyle B^\prime} \\
                S_{\scriptscriptstyle M}&L_{\scriptscriptstyle M} &
                              J_{\scriptscriptstyle M}  \\
                              S_{\scriptscriptstyle B^\prime 
M}&L_{\scriptscriptstyle B^\prime M}&J_{\scriptscriptstyle B^\prime M} 
\end{bmatrix}\nonumber \\
&&\times \sum_L {\hat L}\/^2
 \begin{Bmatrix}s_{\scriptscriptstyle B}&\ell_{\scriptscriptstyle
                              B}&J_{\scriptscriptstyle B} \\
                L&S_{\scriptscriptstyle B^\prime M}&1 \end{Bmatrix}
\begin{Bmatrix}S_{\scriptscriptstyle B^\prime M}&L_{\scriptscriptstyle
                              B^\prime M}&J_{\scriptscriptstyle
                 B^\prime M} \\
                 \ell&J_{\scriptscriptstyle B}&L \end{Bmatrix} 
\varepsilon(\ell_{\scriptscriptstyle B^\prime},L_{\scriptscriptstyle 
M},L_{\scriptscriptstyle B^\prime M},\ell,\ell_{\scriptscriptstyle B},L,k_0).
\end{eqnarray}
\end{widetext}
Here 
\begin{equation}
{\bf J}_B={\bf L}_B + {\bf S}_B = {\bf \ell}_B + {\bf s}_B,
\end{equation}
with
\begin{eqnarray} \label{recoup1}
{\bf L}_B&=&{\bf L}_{\lambda_B} + {\bf L}_{\rho_B}\equiv{\bf \ell}_B + {\bf 
L}_{\rho_B}, \nonumber \\
{\bf S}_B&=&{\bf S}_{\rho_B} + {\bf 1/2},
\end{eqnarray}
and
\begin{eqnarray} \label{recoup2}
{\bf s}_B&=&{\bf J}_{\rho_B} + {\bf 1/2}={\bf L}_{\rho_B} + {\bf S}_{\rho_B} + 
{\bf 1/2},
\end{eqnarray}
with similar definitions for $B^\prime$. The first four $6-j$ symbols
of Eq.~(\ref{ampa1}) are necessary for transforming from the usual
angular momentum basis for the baryons, given by Eq.~(\ref{recoup1}),
to the basis of Eq.~(\ref{recoup2}), which is the more convenient one
for evaluating the transition amplitude. Here $L$, $L_{B^\prime M}$ and 
$S_{B^\prime M}$
are internal summation variables, ${\cal F}(BB^\prime M)$ is the flavor
overlap for the decay, and ${\cal R}(BB^\prime M)$ is the overlap of the wave
functions in the $\rho$ coordinates in the initial and final
baryons. 

The purely ``spatial'' part of the transition amplitude is
\begin{widetext}
\begin{eqnarray} \label{ampa2}
\varepsilon (\ell_{B^\prime},L_M,L_{B^\prime M},\ell,\ell_B,L,k_0)&=&{\cal 
J}(B)(-1)^{L_{B^\prime M}} {1 \over 2}
{\exp{(-F^2k_0^2)} \over G^{\ell_B+\ell_{B^\prime}+L_M+4}} N_B N_{B^\prime} 
N_M\nonumber \\
&&\times \sum_{\ell_1,\ell_2,\ell_3,\ell_4} C^{\ell_{B^\prime}}_{\ell_1} 
C^{L_M}_{\ell_2}
C^1_{\ell_3} C^{\ell_B}_{\ell_4}\left(x-\omega_1\right)^{\ell_1}
\left(x-\omega_2\right)^{\ell_2} \left(x-1\right)^{\ell_3} 
x^{\ell_4}\nonumber \\
&&\times \sum_{\ell_{12},\ell_5, \ell_6, \ell_7, \ell_8}
(-1)^{\ell_{12}+\ell_6} 
{\hat \ell_5 \over \hat
L}\begin{bmatrix}\ell_1&\ell_1^\prime&\ell_{B^\prime} \\
                           \ell_2&\ell_2^\prime&L_M \\
                           \ell_{12}&\ell_6&L_{B^\prime M} \end{bmatrix}
\begin{bmatrix}\ell_3&\ell_3^\prime&1 \\
               \ell_4&\ell_4^\prime&\ell_B \\
               \ell_7&\ell_8&L \end{bmatrix}\nonumber \\
&&\times\begin{Bmatrix}\ell&\ell_{12}&\ell_5 \\
               \ell_6&L&L_{B^\prime M} \end{Bmatrix} B^{\ell_{12}}_{\ell_1 
\ell_2} B^{\ell_5}_{\ell \ell_{12}}
B^{\ell_6}_{\ell_1^\prime \ell_2^\prime} B^{\ell_7}_{\ell_3 \ell_4}
B^{\ell_8}_{\ell_3^\prime \ell_4^\prime}\nonumber \\
&&\sum_{\lambda,\mu,\nu} D_{\lambda \mu \nu}(\omega_1,\omega_2,x) 
I_\nu(\ell_5,\ell_6,\ell_7,\ell_8;L)
\left({\ell_1^\prime+\ell_2^\prime+\ell_3^\prime+\ell_4^\prime+2\mu+\nu+1 
\over 2}\right)!\nonumber \\
&&\times k_0^{\ell_1+\ell_2+\ell_3+\ell_4+2\lambda+\nu}/G^{2\mu+\nu-\ell_1-\ell
_2-\ell_3-\ell_4}.
\end{eqnarray}
\end{widetext}
In this expression, ${\cal J}(B)$ is the Jacobian factor, and $N_B$ is
a normalization coefficient for the wave function of initial baryon
$B$. 

The term $\sum_{\lambda,\mu,\nu} D_{\lambda \mu \nu}(\omega_1,\omega_2,x)
I_\nu(\ell_5,\ell_6,\ell_7,\ell_8;L)$ arises from writing the
product of the associated Laguerre polynomials and exponentials of the
hadron wave functions (here ${\bf q}_B\equiv{\bf p}_{\lambda_B}$, with
a similar definition for the daughter baryon)
\begin{eqnarray}
L_{n_{\lambda_B}}^{\ell_B}e^{-A^2q_B^2/2} L_{n_{\lambda_{B^\prime}}}
^{\ell_{B^\prime}}e^{-B^2q_{B^\prime}^2/2}
L_{n_M}^{L_M}e^{-C^2q_M^2/2}
\nonumber \\
\equiv\sum_{\lambda,\mu,\nu} D_{\lambda \mu
\nu}(\omega_1,\omega_2,x) e^{-A^2q_B^2/2} e^{-B^2q_{B^\prime}^2/2}
e^{-C^2q_M^2/2}. \nonumber \\
\end{eqnarray}
When the substitutions ${\bf q}_B=x{\bf k} +{\bf q}$, ${\bf
q}_{B^\prime}=(x-\omega_1){\bf k} +{\bf q}$, ${\bf
q}_M=(x-\omega_2){\bf k} +{\bf q}$ are made, and the integrals over
${\bf k}$ (the momentum of the final baryon) and ${\bf q}$ are
evaluated, the expression above results with $I_\nu$ a purely
geometric factor.

In Eqs.~(\ref{ampa1}) and (\ref{ampa2}),
\begin{equation}
\begin{bmatrix}a&b&c\\
                 d&e&f\\
                 g&h&i \end{bmatrix}
= \hat c \hat f \hat g \hat h \hat i \begin{Bmatrix}a&b&c\cr
                 d&e&f\\
                 g&h&i \end{Bmatrix}
\end{equation}
where
\begin{equation*}
\begin{Bmatrix}a&b&c\cr
                 d&e&f\\
                 g&h&i \end{Bmatrix}
\end{equation*}
is the $9-j$ symbol, and ${\hat J}\/= \sqrt{2J+1}$.

In Eq.~(\ref{ampa2})
\begin{eqnarray}
x&=&\left(B^2\omega_1+C^2\omega_2+f^2 \right)
\left(A^2 +B^2 +C^2 +f^2 \right)^{-1},\nonumber \\
F^2&=&{1 \over 2}\left[A^2x^2 +B^2\left(x-\omega_1\right)^2+C^2\left(x-\omega_2
\right)^2+f^2(x-1)^2\right],\nonumber \\
G^2&=&{1 \over 2}(A^2+B^2+C^2+f^2).
\end{eqnarray} \\
$\omega_1$ and $\omega_2$ are ratios of various linear combinations of
quark masses. In addition,
\begin{eqnarray}
C^\ell_{\ell_1}&=&\sqrt{{4\pi (2\ell+1)! \over (2\ell_1+1)!
[2(\ell-\ell_1)+1]!}},\nonumber \\
B^\ell_{\ell_1 \ell_2} &=&{(-1)^\ell \over \sqrt{4\pi}} {\hat \ell}\/_1
{\hat \ell}\/_2 \begin{pmatrix}\ell_1&\ell_2&\ell\cr
                 0&0&0 \end{pmatrix},
\end{eqnarray}
and $\ell_1^\prime=L_{B^\prime}-\ell_1$, $\ell_2^\prime=L_M-\ell_2$,
$\ell_3^\prime=1-\ell_3$, $\ell_4^\prime=L_B-\ell_4$.

%
%

%
\end{document}

%% file: decay_fig.tex
\thicklines 
\unitlength 0.575cm
\begin{picture}(16,14.5)
\multiput(2.0,7.0)(0.0,1.0){3}{\line(1,0){5}} 
\multiput(7.0,7.0)(0.0,1.0){2}{\line(3,-1){5.0}} 
\put(2.0,7.2){$1$}
\put(2.0,8.2){$2$}
\put(2.0,9.2){$3$}
\put(12.0,10.3){$3$}
\put(12.0,9.3){$\bar 4$}
\put(12.0,5.3){$1$}
\put(12.0,6.4){$2$}
\put(12.0,7.4){$4$}
\put(7.0,9.0){\line(3,1){5.0}} 
\put(8.0,8.5){\line(3,1){4.0}} 
\put(8.0,8.5){\line(3,-1){4.0}} 
\put(1.2,11.5){\framebox(6.2,1.55){\shortstack{$B:{\bf s}_{\scriptscriptstyle
B}={\bf J}_{\rho_{\scriptscriptstyle B}}+{\bf 
1/2};$\\ {}\\ ${\bf J}_{\scriptscriptstyle B}={\bf s}_{\scriptscriptstyle B}+{\bf L}_{\lambda_{\scriptscriptstyle B}}$}}} 
\put(10.0,3.0){\framebox(6.2,1.6){\shortstack{$B':{\bf s}_{{\scriptscriptstyle B}'}={\bf J}_{\rho_{{\scriptscriptstyle B}'}}+{\bf 
1/2};$\\ {}\\ ${\bf J}_{{\scriptscriptstyle B}'}={\bf s}_{{\scriptscriptstyle B}'}+{\bf L}_{\lambda_{{\scriptscriptstyle B}'}}$}}} 
\put(10.0,11.5){\framebox(6.2,1.55){\shortstack{$M:{\bf S}_{\scriptscriptstyle M}={\bf 1/2}+{\bf 
1/2};$\\ {}\\ ${\bf J}_{\scriptscriptstyle M}={\bf S}_{\scriptscriptstyle M}+{\bf L}_{\scriptscriptstyle M}$}}} 
\put(10.0,0.5){\framebox(6.2,1.55){\shortstack{$B'M:{\bf J}_{\scriptscriptstyle
{B'M}}={\bf J}_{\scriptscriptstyle B'} + 
{\bf J}_{\scriptscriptstyle M};$\\ {}\\ ${\bf J}_{\scriptscriptstyle B}={\bf J}_{{\scriptscriptstyle B'M}}+{\bf \ell}$}}} 
\end{picture} 

%% file: paper.bbl
\begin{references}
\bibitem{dressed-bare} In field theory this would correspond to using
dressed propagators for the intermediate particles and so working in
rainbow approximation.
\bibitem{whyf=0?} It was found in Ref.~\cite{CR} that the value of
$f$ is not tightly constrained by baryon pionic decays, and a fit of
acceptable quality is obtained with $f=0$. Since hadron loops
sample all relative momenta of the final state hadrons, they tightly
constrain the value of $f$.
\bibitem{numloops} For a general external nucleon state 591 loops
contribute, and for $\Delta$ state 378 loops are required.
\end{references}
